\documentclass[10pt,conference]{IEEEtran}
\usepackage[english]{babel}
\usepackage[font=small,labelfont=bf]{caption}
\usepackage{amsmath}
\usepackage{ifthen}
\usepackage{amsthm}
\usepackage{amsfonts}
\usepackage{epsfig}
\usepackage[table]{xcolor} 
\usepackage{graphics,color} % for pdf, bitmapped graphics files
\usepackage{rotating}
\usepackage{algpseudocode}
\usepackage{graphicx}
\usepackage{comment}
\usepackage{multirow}
\usepackage{xcolor}
\usepackage{algorithm}
\usepackage{url}
\usepackage{epsfig}
\usepackage{url}
\usepackage{cases}
\usepackage{mathtools}
\usepackage{amssymb} % for \mathbb
\usepackage{dsfont}
\usepackage{cleveref}
\usepackage{bbm}
\usepackage{dblfloatfix}
\usepackage{tikz} % For tikz figures (to draw arrow diagrams, see a guide how to use them)
\usepackage{tikz-cd}
\usetikzlibrary{positioning,arrows} % Adding libraries for arrows
\usetikzlibrary{decorations.pathreplacing} % Adding libraries for decorations and paths
\usepackage{tikzsymbols} % For amazing symbols ;) https://mirror.hmc.edu/ctan/graphics/pgf/contrib/tikzsymbols/tikzsymbols.pdf 
\usepackage{blindtext} % To add some blind text in your paper
\usepackage{tikz}
\usetikzlibrary{fit,tikzmark}
\newtheorem{lemma}{Lemma}

\newtheorem{theorem}{Theorem}
\newtheorem{corollary}{Corollary}
\newtheorem{proposition}{Proposition}
 
\newtheorem{conjecture}{Conjecture}

\crefname{lemma}{Lemma}{Lemmas}
\crefname{definition}{Definition}{Definitions}
\crefname{theorem}{Theorem}{Theorems}
\crefname{corollary}{Corollary}{Corollaries}
\crefname{proposition}{Proposition}{Propositions}
\crefname{remark}{Remark}{Remarks}
\crefname{conjecture}{Conjecture}{Conjectures}
\crefname{algorithm}{Algorithm}{Algorithms}
\crefname{assumption}{Assumption}{Assumptions}

\newcommand{\rp}[1]{\left( #1 \right)}
\newcommand{\A}{\mathcal{A}}

\renewcommand{\S}{\mathcal{S}}
\newcommand{\E}[1]{\mathbb E  \left[ #1 \right]}
\newcommand{\expvalDist}[2]{\mathbb{E}_{#1} \left[ #2 \right]}

\newcommand{\Pro}[1]{\mathbb{P} \rp{#1} }		% p
\newcommand{\tM}{\widetilde{\mathcal{M}}}
\newcommand{\M}{\mathcal{M}}

\newcommand {\R}{\mathbb{R}}	

\newcommand{\1}[1]{\mathbbm{1}\!\!\left \{#1\right \}}	
\newif\ifextver                             % enables the extended version 
\extvertrue                                 % it is the extended version
\IEEEoverridecommandlockouts 
\begin{document}

\title{Performing Load Balancing under Constraints}
\author{Andrea Fox$^{1}$, Francesco De Pellegrini$^{1}$, Eitan Altman$^{2}$, Arnob Ghosh$^{3}$, and Ness Shroff$^{4}$\thanks{$^{1}$LIA, Avignon university, Avignon, France; $^{2}$INRIA, Sophia Antipolis, France; $^{3}$New Jersey Institute of Technology, Newark, NJ, USA; $^{4}$The Ohio State University, Columbus, OH, USA;}}
%\author{}
\maketitle
\begin{abstract}
Join-the-shortest queue (JSQ) and its variants have often been used in solving load balancing problems. The aim of such policies is to minimize the average system occupation, e.g., the customer's system time. In this work we extend the traditional load balancing setting to include constraints that may be imposed, e.g., due to the communication network. We cast the problem into the framework of constrained MDPs, enabling the consideration of both action-dependent constraints, such as, e.g, bandwidth limitation, and state-dependent constraints, such as, e.g., minimum queue utilization. Unlike the state-of-the-art approaches, our load-balancing policies, in particular JSED-$k$ and JSSQ, are both provably safe and yet strive to minimize the system occupancy. Their performance is tested with extensive numerical results under various system settings.
\end{abstract}

%%%%%%%%%%%%%%%%%%%%%%%%%%%%%%%%%%%%%%%%%%%%%%%%%%%%%%%%%%%%%%%%%%%%%%%%%%%%%%%%%%%%%%%%%%%
%%%%%%%%%%%%%%%%%%%%%%%%%%%%%%%%%%%%%%%%%%%%%%%

\section{Introduction}

%%%%%%%%%%%%%%%%%%%%%%%%%%%%%%%%%%%%%%%%%%%%%%%
%%%%%%%%%%%%%%%%%%%%%%%%%%%%%%%%%%%%%%%%%%%%%%%%%%%%%%%%%%%%%%%%%%%%%%%%%%%%%%%%%%%%%%%%%%%
Load balancing is a common strategy for task assignment in distributed architectures, including web services \cite{gupta2007analysis}, distributed caching systems \cite{nishtala2013scaling}, and cloud computing \cite{foster2008cloud}. In these systems, a job dispatcher seeks to distribute incoming jobs across servers as they enter the system, aiming to minimize the customers' queueing delays.

The performance of a load balancing system largely depends on the policy employed. A markovian policy, as those studied in this paper, decides the queue where to dispatch an incoming arrival depending on the current system state. In the literature on queueing theory, several works have tackled this problem using different approaches, including \cite{winston1977optimality, weber1978optimal,stolyar2005optimal}.

Despite the extensive body of research traditionally covered in the literature on load balancing, to the best of the authors' knowledge, the problem has not been studied so far \textit{in constrained systems}. Indeed, many real-world applications could benefit from addressing this facet of the problem. Hence, in this paper our objective is to design new policies able to provide solutions for constrained load balancing, i.e., safe load-balancing policies. In particular, we restrict to two scenarios for constrained load balancing and define the related constraints accordingly. 

%A similar approach has been employed in video streaming for high-quality virtual reality applications \cite{nguyen2018VRvideoStreaming}. Moreover, the integration of constant video quality rate control into the state-of-the-art H.264/AVC video encoder, as demonstrated in \cite{ozcelebi2006analysis}, has shown strong performance, especially in scenarios with high traffic demands.
% we assume that each job execution consumes a fixed fraction of the available channel capacity. %of the link connecting the central dispatcher to each queueand the other is based on a generic cost function that depends on the occupancy of each queue.
%The first type of constraint is a communication constraint. It occurs because exceeding the control channel capacity between the load balancer and servers can lead to latency and packet loss \cite{al2012simulation}. Similarly, in IoT systems, exceeding the bandwidth for offloading jobs to edge servers may disrupt communication and task allocation. Ignoring channel capacity constraints risks congestion, degraded performance, or operational failures. Our study provides robust, efficient load-balancing algorithms suited for practical conditions and real-world scenarios.
The first type of constraint is a communication one. Jobs are uniform in their resource demand: each job may be different in service time, but it consumes a fixed fraction of resources, i.e., of the available channel capacity. Such type of constraint may represent, e.g., the finite capacity of the control channel connecting a load balancer and its servers. Violating such constraints can lead to higher latency and packet loss \cite{al2012simulation}. Similarly, in IoT systems, exceeding the access bandwidth could disrupt communication and task allocation when executing jobs offloaded to edge servers. Also, in content delivery networks, multiple users request different video streams and each stream consumes an average fixed portion of the available bandwidth, as done e.g., in video streaming  rate control \cite{ozcelebi2006analysis} and virtual reality applications \cite{nguyen2018VRvideoStreaming}. 

The second type of constraint we consider is meant to ensure a minimum level of activity on each queue to prevent under-utilization, particularly in low-traffic scenarios. %In practice, the operational costs of maintaining physical resources such as machines, or facilities suggest to avoid the presence of idle servers. 
In practice, the high operational costs of maintaining physical resources like, for instance, machines and facilities in a datacenter, impose to avoid idle servers \cite{xu2015traffic}. This becomes particularly critical in systems subject to rapid traffic load fluctuations. In fact, dynamic server provisioning is usually unable to effectively respond to small-time-scale traffic variations due to the substantial time delay introduced by powering servers on and off. 

%This ensures that the system operates in a balanced and efficient manner, avoiding scenarios where certain resources remain idle or underused. 
% In addition, by distributing demand more evenly across the queues prevents over-reliance on specific ones while ensuring all resources are actively contributing to the system’s performance.
% Ultimately, these constraints align resource usage with demand, ensuring cost-effectiveness and a consistent level of performance while enhancing user experience and satisfaction. 
%This framework supports not only the efficient operation of the queueing system but also its long-term sustainability and effectiveness.
In this work, we propose a Constrained Markov Decision Process (CMPD) framework to address the problem of constrained load balancing. The model can incorporate the types of constraints described before and it can be applied flexibly to different systems. We also describe the structure of a policy that serves as a close approximation of the optimal solution, although we believe it remains impractical for real-world applications due to its computational complexity. Finally, we study new lightweight policies which are provably safe.%, i.e., they satisfy the constraints. %They are designed with specific properties to accommodate varying traffic load requirements. They can hence offer practical and flexible solutions for constrained load balancing across various operational scenarios. 

%Specifically, one policy utilizes virtual queues to model the link capacity constraints, another maintains a memory of the past $k$ actions to assess the satisfaction of every constraint, and the final policy solves a nonlinear problem related to the Markov Decision Process (MDP) to determine a feasible solution for the arrival rates at each server.  The numerical results demonstrate that the solution which stores the last $k$ actions in memory performs well under lighter traffic conditions but may become impractical in heavier traffic scenarios. In contrast, the alternative approach consistently delivers a policy that satisfies all the constraints across all tested systems.
Specifically, the first policy, namely JSVED, relies on virtual queues to model the link capacity constraints and attains a conservative and yet safe solution. The second policy, namely JSED-$k$, maintains a memory of the last $k$ actions to assess the satisfaction of the constraints. The numerical results demonstrate that JSED-$k$ performs well under lighter traffic conditions but may become impractical in heavier traffic scenarios. For this reason, we propose a third alternative policy, namely JSSQ,  which is determined by solving first a convex nonlinear problem related to the CMDP, whose solution provides a vector of feasible arrival rates at each server. This overcomes scalability issues at the price of larger system occupation and provably satisfies the constraints.

The paper is structured as follows. Section \ref{sec:sysmod} introduces our reference model. Section \ref{sec:optimal policy properties} describes key properties of the optimal policy and presents a near-optimal approach via the Lyapunov-drift method. Section \ref{sec:safe policies} discusses new lightweight approximation policies that satisfy the constraints, while Section \ref{sec:generic cost constraints} addresses the problem using a generic gain function based on the occupancy of each queue. Section \ref{sec:related works} reviews related works in the literature. Simulation results are presented in Section \ref{sec:numerical results}. A final section concludes the paper.

%%%%%%%%%%%%%%%%%%%%%%%%%%%%%%%%%%%%%%%%%%%%%%%%%%%%%%%%%%%%%%%%%%%%%%%%%%%%%%%%%%%%%%%%%%%
%%%%%%%%%%%%%%%%%%%%%%%%%%%%%%%%%%%%%%%%%%%%%%%

\section{System Model}\label{sec:sysmod}

%%%%%%%%%%%%%%%%%%%%%%%%%%%%%%%%%%%%%%%%%%%%%%%
%%%%%%%%%%%%%%%%%%%%%%%%%%%%%%%%%%%%%%%%%%%%%%%%%%%%%%%%%%%%%%%%%%%%%%%%%%%%%%%%%%%%%%%%%%%

%%%%%%%%%%%%%%%%%%%%%%%%%%%%%%%%%%%%%%%%%%%%%%%%%%%%%%%%%%%%%%%%%%%
\begin{table}[t!]
	\centering
	\begin{tabular}{|p{0.16\columnwidth}|p{0.68\columnwidth}|}
		\hline
		\rowcolor{gray!30}{\bf Symbol} & {\bf Meaning}                         \\
		\hline
		  $N$          & number of queues                      \\
        $M$          & number of constrained queues constrained \\
        $\Xi$    & arrival rate of the system       \\
        $\xi_i$ & arrival rate of queue $i$ \\
        $\mu_i$      & service rate of queue $i$      \\
        $\theta_i$   & constraint for queue $i$ \\
        $\mu_i'$ & capacity of virtual queue $i$ -- link capacity constraints. 
        $\quad \rp{ \mu_i' = \min(\mu_i, \theta_i) }$ \\
		\hline
	\end{tabular}\caption{Main model notation.}\label{tab:notation}
\end{table}
%%%%%%%%%%%%%%%%%%%%%%%%%%%%%%%%%%%%%%%%%%%%%%%%%%%%%%%%%%%%%%%%%%%

\begin{figure}[t]
    \centering
    \includegraphics[width = 0.8\linewidth]{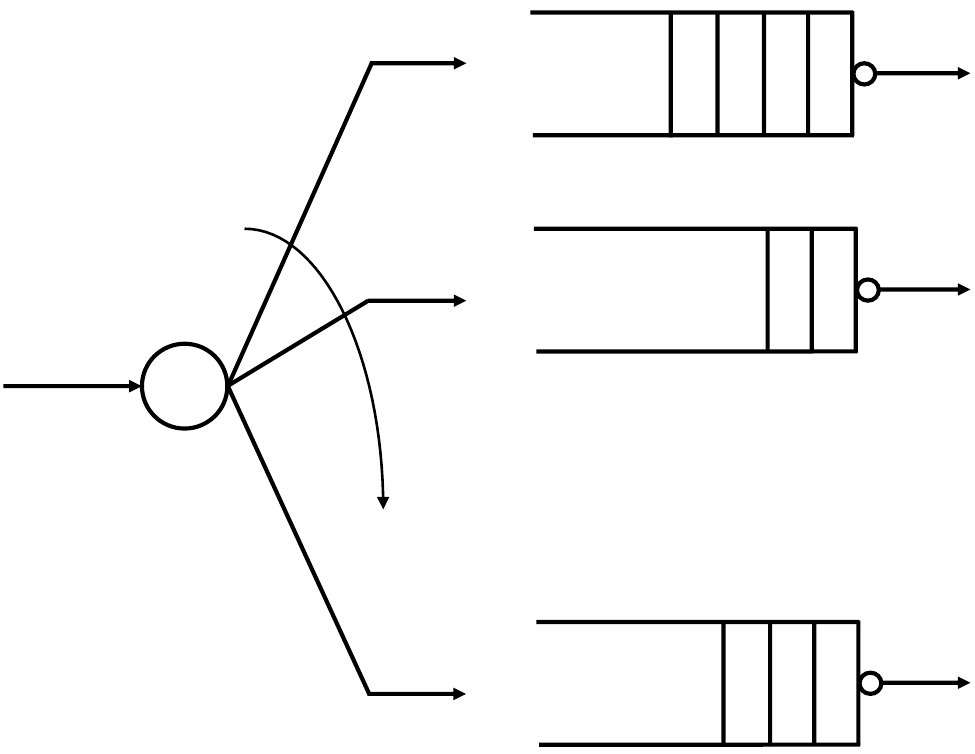}
    \put(-170,72.5){{\small $LB$}} \put(-192,80){$\lambda$}\put(-130,40){$A_t \sim \pi $}
    \put(-16,20){$\mu_{_N}$}\put(-16,103){$\mu_{_2}$}\put(-16,148){$\mu_{_1}$}
    \put(-150,33){$\theta_{_N}$}\put(-130,103){$\theta_{_2}$}\put(-150,123){$\theta_{_1}$}
    \caption{Load balancing for a parallel of $N$ queues with access constraint.}%
    \label{fig:loadbalancing}
\end{figure}
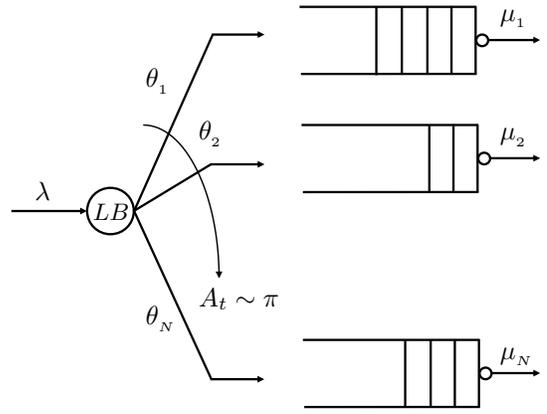

We consider a system of $N$ parallel servers or queues as depicted in \cref{fig:loadbalancing}. Server $1\leq i\leq N$ has an exponential service time with known rate $\mu_i$, and all service times are independent. Clients arrive according to a Poisson process of intensity $\Xi$ and a load balancer immediately dispatches each customer to a given server. 

Throughout this work, uppercase notation (e.g., $S$) represents a random process, while lowercase notation (e.g., $s$) refers to its realization. We consider the uniformization of the continuous-time process, resulting in the discrete-time Markov Decision Process (MDP) $\M = (\S, \A, P)$ \cite{lippman1975applying}. The state space $\S$, action set $\A$, and transition probabilities are defined hereafter. 

The state space $\S$ consists of nonnegative integer vectors $s = (s_1, \ldots, s_N)$, where $S_t = s$ represents the system state at time $t$. Here, $s_i$ denotes the occupation of server $i$ for $i = 1, \dots, N$. It is important to observe that sampling the queues occupation may becomes increasingly challenging as the number of queues grows. In this work, we assume that the associated overhead remains manageable, even under heavy-traffic conditions. Methods that address this issue in the context of unconstrained load balancing have been explored in \cite{zhouAsymptoticallyOptimalLoad2021,zhouHeavytrafficDelayOptimality2018a,zhouDegreeQueueImbalance2018,zhouDesigningLowComplexityHeavyTraffic2017}. Extending these approaches to constrained load balancing scenarios is left for future work.

The action set $\A = \{1, 2, \ldots, N\}$ specifies the server a client is dispatched to: $A_t = i$ indicates that the client arriving at time $t$ is assigned to server $i$. Define the transition probabilities $p(s'|s,i) := \Pro{S_{t+1}= s'|S_{t}=s,A_t=i}$ and let $\widehat p(u; s_j)$ denote the probability that $u$ customers of server $j$ leave in between two arrivals, given $s_j$ customers are present at the beginning of the inter-arrival interval.
\begin{equation}\label{eq:deathprob}
	\widehat p(u; s_j)= \Xi \int_0^\infty \binom{s_j}{u} e^{-\mu_j t(s_j- u)} (1-e^{-\mu_j t})^u e^{-\Xi t} dt \nonumber
\end{equation}
for $0 \leq u \leq s_j$ and zero otherwise. Hence, the transition probabilities can be written as
\begin{align}
    p(s'|s,i) = \widehat p (s_i - s_i'+1; s_i) \prod_{j \not = i}\widehat p (s_j - s_j'; s_j)
    \label{eq:transition probabilities}
\end{align}
With standard notation, $R_{t+1}$ is the reward attained after the action taken at time $t$ and $r(s,i)=\E{R_{t+1}|S_t=s,A_t=i}$. We define the immediate reward at time $t$ as the time to serve all the customers currently in service, so that
\begin{equation}
r(s,i) =  \frac {s_i+1}{\mu_i} + \sum\limits_{j \not = i}  \frac {s_j}{\mu_j}
    \label{eq:immediate reward}
\end{equation}
We consider the set of randomized stationary policies $\Pi$, where policy $\pi$ is a probability distribution over the state-action space set. In an unconstrained setting, the admission control problem can be written as 
$$\mathop{\mbox{minimize:}}_{\pi \in \Pi} \ \expvalDist{\pi, s_0}{\lim_{T\to\infty} \frac{1}{T} \sum_{t=0}^T r\rp{S_t, \pi(S_t)} }$$
%the objective function to maximize for the admission control problem is the average expected reward %$G_t:= \sum_{t=0}^{\infty} {R_{t+1+k}}$ 
for an initial state $s_0$. For every stationary deterministic policy, the resulting Markov chain is regular, meaning it has no transient states and a single recurrent acyclic class. 
We define the value function
\begin{align*}
v_\pi(s) =\lim_{T \to \infty} \frac{1}{T}\sum\limits_{t=0}^T \expvalDist{\pi}{r \rp{S_t,A_t}| S_0=s} % \label{eq:avcost2}
\end{align*}
%Due to the presence of constraints \cite{altman2021constrained} 
Finally, the average reward is taken w.r.t. the initial state distribution $\beta:\S \rightarrow \Delta$ \cite{altman2021constrained} 
\begin{equation}\label{eq:avcost1}
	J_\pi(\beta)= \expvalDist{S\sim \beta}{v_\pi(S)}
\end{equation}
Next, we introduce the CMDP formulation to account for the constraints on the queue access capacity.

%%%%%%%%%%%%%%%%%%%%%%%%%%%%%%%%%%%%%%%%%%%%

\subsection{The CMDP model}\label{subsec:cmdp}

%%%%%%%%%%%%%%%%%%%%%%%%%%%%%%%%%%%%%%%%%%%%
For now, we focus on the case where the constraint represents the channel capacity constraints, e.g., the capacity of the link connecting each server to the central dispatcher. The second case of constraint, %more general scenario, 
involving a gain function dependent on the occupation of each server will be analyzed in Section \ref{sec:generic cost constraints}. The constraint will be represented by the vector $\theta \in \mathbb{R}^M$ and can be interpreted as the fraction of the arrivals that can be routed towards each constrained queue. The instantaneous cost $c_i: S\times A\rightarrow \R$ related to the access to queue $i$ is independent of the state of the queue and it is defined as: 
\begin{equation}
c_j(s,i) = \1{j=i} 
\label{eq:immediate cost}
\end{equation}
% In future work, it may be beneficial to consider a scenario where each arrival has a different weight, allowing for a more flexible and realistic adaptation to varying system conditions.
The vector $K_\pi(\beta)=(K_\pi^1(\beta),\ldots,K_\pi^N(\beta))$ represents the average cumulative constraint:
\begin{equation}\label{eq:constraint}
	K_{\pi}^i(\beta)= \expvalDist{\pi, s\sim \beta}{  \lim_{T \to \infty} \frac{1}{T}\sum\limits_{t=0}^{T}  c_i \big( S_t,A_t \big) | S(0)=s}  
\end{equation}
For a fixed access capacity vector $\theta=( \theta_1,\ldots, \theta_N)$, and a feasible initial state distribution $\beta$, we seek an optimal policy solving the constrained load balancing control  (\ref{eq:CLBC}) problem
\begin{align}
	\mathop{\mbox{minimize:}}_{\pi \in \Pi}   \; J_\pi&(\beta)  \label{eq:CLBC}\tag{CLBC}       \\
	\mbox{subject to:} \;                           K_\pi&(\beta) \leq  \theta \label{eq:constr}
\end{align}
We let $J^*(\beta)$ denote the corresponding optimal value. 

Throughout the paper, we assume that queues $1, \dots, M$ (M could be equal to N) are those constrained by their link capacities, while the remaining $N-M$ queues have no link capacity constraint.

\begin{comment}
Let us define $\rho(s,i)$ the stationary state-action distribution. From a know result in CMDP theory \cite{altman2021constrained}  
\begin{lemma}\label{lem:dualLP}
An optimal stationary state-action distribution $\rho^*$ for \ref{eq:CLBC} solves the following dual linear program
\begin{align}
\mathop{\mbox{maximize:}}_{\rho(s,i)}   \; & \sum_{s,i}r(s,i) \rho(s,i)   \label{DLP}\tag{DLP}                                                                                                       \\
\mbox{subj. to:}                 & \sum_{s,i} \rho(s,i) [ \delta_{s\underline s}- \gamma p(\underline s |s,i)] = \beta(\underline s), \; \underline s\in \S \label{eq:bellman}             \\
	                                     & \sum_{s,i} \1{j=i}  \rho(s,i)\leq  \theta_i, \; 1\leq i\leq N  \label{eq:constraintApp}                                                        \\
	                                     & \rho(s,i)\geq 0, \quad \forall (s,i)  \in \S \times \A  \label{eq:nonnegApp}
\end{align}
where $\beta$ is the initial distribution and $\delta_{s\underline s}=1$ if $s=\underline s$ and zero otherwise. The corresponding optimal policy writes, for non transient states, as  
\begin{equation}
\pi^*(a|s)=\frac{\rho^*(s,a)}{\sum_{a'\in \A} \rho^*(s,a')}
\end{equation}
\end{lemma}
\end{comment}
%%%%%%%%%%%%%%%%%%%%%%%%%%%%%%%%%%%%%%%%%%%%%%%%%%%%%%%%%%%%
\subsection{Condition for the stability of the system}
%%%%%%%%%%%%%%%%%%%%%%%%%%%%%%%%%%%%%%%%%%%%%%%%%%%%%%%%%%%%
In the absence of constraints, if $\Xi < \sum_i \mu_i$, then there is a policy that stabilizes the system, that is, such that $\sum_{s\in \S} \Xi \pi(i|s) \rho(s)< \mu_i$ for all $i$s; for example, the trivial constant policy $\pi(i|.)= \min\{\mu_i,\frac{\mu_i}{\sum_j \mu_j}\}$ does so.

Under the presence of constraints, we define an equivalent Markov Decision Process (MDP) by modifying the service rates $\mu_i$. Specifically, for the constrained queues ($1, \dots, M$), the service rates are adjusted to $\mu_i' = \min \{ \Xi \theta_i, \mu_i\}$, while for the remaining queues, the service rates remain unchanged, i.e., $\mu_i' = \mu_i$. In this formulation, $\mu_i'$ for the constrained queues represents the maximum arrival rate that the link can handle under the given constraints. Notably, for these queues, $\mu_i'$ is smaller than their original processing capacity $\mu_i$, reflecting the imposed limitations. % Here, $\mu_i'$ represents the new maximum processing capacity of the $i$-th queue, determined by the lesser of the queue's processing speed $\mu_i$ and the capacity of the link $\Xi \theta_i$. 
In this context, we have actual queues replaced by virtual ones. Stability holds if $\Xi < \sum_i \mu_i'$.  
To stabilize the system under these constraints, we can construct a policy that allocates arrivals proportionally to the modified capacities. %Specifically, the policy is defined as $$\pi'(i \mid \cdot) = \frac{\mu_i'}{\sum_j \mu_j'}$$
This ensures that the system operates within the constraints while maintaining stability.

%%%%%%%%%%%%%%%%%%%%%%%%%%%%%%%%%%%%%%%%%%%%%%%%%%%%%%%%%%%%%%%%%%%%%%%%%%%%%%%%%%%%%%%%%%%%%%%%%%%%%%%%%%%%%%%%
%%%%%%%%%%%%%%%%%%%%%%%%%%%%%%%%%%%%%%%%%%%%%%%%
\section{Optimal policy properties}
\label{sec:optimal policy properties}
%%%%%%%%%%%%%%%%%%%%%%%%%%%%%%%%%%%%%%%%%%%%%%%%
%%%%%%%%%%%%%%%%%%%%%%%%%%%%%%%%%%%%%%%%%%%%%%%%%%%%%%%%%%%%%%%%%%%%%%%%%%%%%%%%%%%%%%%%%%%%%%%%%%%%%%%%%%%%%%%%

The optimal policy in the \textit{unconstrained} case for a setting with homogeneous servers is {\em Join the Shortest Queue} (JSQ). Following intuition, this policy assigns each new arrival to the queue with the shortest current length. The optimality of JSQ was first established in \cite{winston1977optimality}. 

In this work, however, we focus on the case of heterogeneous queues, where servers may have different processing speeds. In the \textit{unconstrained} case, the optimal policy is JSED (\textit{Join the Shortest Expected Delay}), that generalizes JSQ by routing arrivals to the queue with the smallest expected response time. 
The asymptotic optimality of JSED under heavy-traffic conditions was proved in \cite{mandelbaum2004scheduling,foschini1977heavy}. 

Let $s_i$ denotes the current number of jobs in queue $i$, and $\mu_i$ represent the processing speed of the $i$-th server. JSED's action is chosen at each time step according to the rule 
\begin{equation} 
    \pi_{\mbox{\tiny JSED}}(s) = \arg\!\min_i \left\{ \frac{s_i}{\mu_i} \right\}
    \label{eq:JSED action}
\end{equation}

Both JSQ and JSED are Markovian policies, which act based on the system state, i.e., the system occupation. Since the problem studied in this work is the optimization of a \textit{constrained} problem, it is natural to resort to CMDP models. From the theory of CMDPs, see Thm 4.4 in  \cite{altman2021constrained}, an optimal policy is stochastic in at most $M$ states and deterministic in all the others. 
\begin{comment}
    Furthermore, with respect to the structure of the optimal policy, we aim to derive a policy that satisfies the following property:
\begin{conjecture}
    Let $\pi^\star : \S \times \A \to [0, 1]$ be the optimal policy, i.e. the one solving \eqref{eq:CLBC}.     If $\pi^\star (s, i) = 1$, then $\pi^\star (s', i) = 1$ whenever $s = \{ x_1, \dots, x_N \}$ and $s' = \{ x_1', \dots, \x_N' \}$ with $s_i' < s_i$. 
    \label{prop:structure optimal policy}
\end{conjecture}
\end{comment}
As proved in Theorem 12.7 in \cite{altman2021constrained}, the solution of \eqref{eq:CLBC} is equivalent to the solution of a MDP whose reward function is
\[
J^\lambda_{eq}(\beta, \pi) = J_{\pi}(\beta) + \langle \lambda, K_{\pi}(\beta) - \theta \rangle
\]
Let us fix the value of the Lagrange multiplier $\lambda$: we can study a new MDP where the immediate reward writes  
\begin{equation}
    r(s, a) = \sum_{i=1}^N \frac{s_i}{\mu_i} + \delta_i (a) \lambda_i
    \label{eq:reward equivalent MDP}
\end{equation}
where $\delta_i (\cdot)$ is the indicating function having value equal to one when the argument is $i$ and $0$ otherwise. 
This leads to a relaxed problem where the objective is to minimize the average reward using the instantaneous penalized reward defined in \eqref{eq:reward equivalent MDP}.

We leave the study of the structure of optimal policis of the CMDLP for future works. For the moment, we address a method which combines a Lyapunov drift-plus-penalty optimization technique and a Lagrangian approach to design a policy that stabilizes the system while attempting to minimize the time-average network occupancy.
%%%%%%%%%%%%%%%%%%%%%%%%%%%%%%%%%%%%%%%%%%%%%%%%%%%%%%%%%%%%
\subsection{Lyapunov optimization}
\label{subsec:Lyapunov optimization LM fixed}
%%%%%%%%%%%%%%%%%%%%%%%%%%%%%%%%%%%%%%%%%%%%%%%%%%%%%%%%%%%%
Lyapunov optimization performs queue stabilization based on a %is a queue stabilization technique based on a  %. It is flexible and effective control technique that stabilizes real or virtual queues while simultaneously optimizing a performance objective. Its appeal 
 greedy, real-time optimization strategy which does not require to know the underlying system dynamics. It is possible to define a parameter $V\geq 0$ to tune the balance between two conflicting objectives: the stability of the system, i.e., of individual queues, and the time average of a certain penalty function. In our case, the penalty is represented by the expected instantaneous reward, namely, the expected makespan corresponding to the current state-action pair $(S_t,A_t)$. We shall leverage a fundamental result tying the time average of the penalty function, which can be rendered %arbitrarily 
 close to its optimal value within $O(1/V)$ \cite{neely2010queue}. %However, this comes at the price of queue backlogs possibly growing as $O(V)$. 

 Lyapunov online optimization can be performed directly while treating the instantaneous reward of the relaxed problem \eqref{eq:reward equivalent MDP} as the instantaneous penalty. %Hence, the reward can still be analyzed using the Lyapunov drift-plus-penalty technique \cite{neely2022stochastic}. 
%In fa, it is possible to perform queue stabilization while minimizing the time-average of a target penalty function. 
We follow the procedure described in \cite{neely2010queue,bracciale2020lyapunov,neely2022stochastic} and apply Theorem 4.2 in \cite{neely2022stochastic}. The resulting greedy policy, at each timestep $t$, performs action 
\[
A_t=\arg\min_i \left \{ V\cdot r(s, a) + \sum_{i=1}^N s_i(t) \rp{\alpha_i(t) - \eta_i(t)} \right \}  
\]
where $\alpha_i$ denotes the arrivals dispatched to queue $i$ and $\eta_i$ the corresponding departures.

In particular, if we make the approximation $\alpha_i(t) - \eta_i(t) \sim 0$ and we impose $V=1$, we obtain the 
following greedy action choice
%If we analyze the equivalent MDP under the penalized reward, since $\alpha_i(t) = \delta_i(a)$, and because the departure process is independent of the action, the resulting greedy action choice is given by
%\begin{equation}
%A_t=\arg\min_a \bigg \{\frac{s_a(t)+1}{\mu_a} + \lambda_a  \bigg \} \qquad \forall t
%    \label{eq:greedy action Lyapunov}
%\end{equation}
%where we have have assumed $V=1$. 
\begin{equation}
A_t=\arg\min_a \bigg \{\frac{s_a(t)+1}{\mu_a} + \lambda_a  \bigg \} \qquad \forall t
    \label{eq:greedy action Lyapunov}
\end{equation}
which in fact corresponds to the minimization of the instantaneous reward of the equivalent MDP. 
% With this choice of the parameter $V$, and applying \cref{thm:Lyapunov optimization}, we can conclude that the penalty averaged in time is at most $O(1)$ above the desired target, while the average queue size remains $O(1)$.
Actually, with this particular choice of parameter $V$, the resulting policy is the Join-the-Minimum-Cost-Queue (JMCQ) policy described in \cite{tandra2004join}. This policy is designed to manage a system of parallel queues offering differentiated service levels to a multiclass customer base. JMCQ operates by assigning different joining costs to each queue, which correspond to the Lagrange multiplier in our framework. 

For the general case, the following result holds: 
\begin{proposition}
    For $\lambda \in \mathbb{R}^M$, $V \in \mathbb{R}$ and let $r^*$ the optimal reward we can obtain in this setting. There exists a greedy load-balancing policy $\tilde{\pi}$ such that time-average of the reward function $\overline{r}$ satisfies $\overline{r} \leq r^\star + O(1/V)$, and $\lim_{T \to \infty} \frac{1}{T} \sum_{t=1}^T \sum_i s_i(T) \leq O(V)$. 
    \label{prop:Lyapunov method}
\end{proposition}
%\label{proof:JSSQ}%\label{proof:JSEDk}%\label{proof:lyap}
The full proof is reported 
\ifextver
in Appendix \ref{proof:JSEDk}.
\else
in \cite{extendedVersion}.
\fi
A further, natural optimization step would require to optimize the multiplier vector $\lambda \in \mathbb{R}^M$, e.g., by using standard stochastic approximation techniques \cite{fu1997optimization}. However, learning optimal Lagrange multipliers through consecutive iterations has two major drawbacks. The first drawback is that it may involve the evaluation of policies that do not satisfy the constraints during the online learning process. At runtime, undesirable outcomes may occur, possibly compromising system reliability or performance. The second one is that this primal-dual method becomes increasingly inefficient as the number of queues grows, since it entails optimization across a high-dimensional space. We defer to \cref{sec:numerical results} for a detailed discussion on the viability of the primal-dual approach in our context. In the next section we focus on a class of lightweight algorithms to solve the constrained load balancing problem. 

%%%%%%%%%%%%%%%%%%%%%%%%%%%%%%%%%%%%%%%%%%%%%%%%%%%%%%%%%%%%%%%%%%%%%%%%%%%%%%%%%%%%%%%%%%%%%%%%%%%%%%%%%%%%%%%%%%%%%%%%
%%%%%%%%%%%%%%%%%%%%%%%%%%%%%%%%%%%%%%%%%%%%%%%%%%%%%%%%%%%%
\section{Finding a safe policy}
\label{sec:safe policies}
%%%%%%%%%%%%%%%%%%%%%%%%%%%%%%%%%%%%%%%%%%%%%%%%%%%%%%%%%%%%
%%%%%%%%%%%%%%%%%%%%%%%%%%%%%%%%%%%%%%%%%%%%%%%%%%%%%%%%%%%%%%%%%%%%%%%%%%%%%%%%%%%%%%%%%%%%%%%%%%%%%%%%%%%%%%%%%%%%%%%%

%As discussed in \cref{sec:optimal policy properties}, the approach based on the optimization of Lagrange multipliers %is not universally applicable. 
%Here, we present strategies for designing safe and robust policy capable of functioning effectively in all scenarios.
Throughout the rest of the paper, we will denote policies that satisfy all the constraints as \textit{safe}.
The priorities in the design of the safe algorithms presented hereafter are {\em scalability}, namely, in the number of served queues, and {\em online safety}, i.e., they should avoid the execution and evaluation of policies violating constraints to grant stability and safety throughout the optimization process.
%%%%%%%%%%%%%%%%%%%%%%%%%%%%%%%%%%%%%%%%%%%%%%%%%%%%%%%%%%%%
\subsection{JSVED: Greedy policy w.r.t. virtual queues}
\label{subsec:JSVED}
%%%%%%%%%%%%%%%%%%%%%%%%%%%%%%%%%%%%%%%%%%%%%%%%%%%%%%%%%%%%
The first solution uses virtual queues \cite{neely2010queue}. We define an auxiliary unconstrained MDP, denoted by $\tM$. In $\tM$ each queue of the original system is associated with an additional virtual queue. Such virtual queue has a processing speed corresponding to the access capacity constraint of its associated real queue, namely $\mu_i' =\theta_i$. In $\tM$, for every new arrival routed to queue $i \in \{1, \dots, M\}$, a corresponding virtual arrival is also routed to the associated virtual queue. The departures of the virtual queue are simulated according to virtual queue processing rate $\mu_i'$. 

%It is trivial to prove that a policy $\pi$ satisfies the constraint in $\M$ if and only if it also ensures the stability of $\tM$, based on the inherent properties of the queues. 

The first policy follows the JSED based on the simulated $\tM$ dynamics. For the constrained queues, it acts based on  the state of the corresponding virtual queues as well as their processing speed. For every constrained queue $i \in \{1, \dots, M \}$, let $\tilde{s}_i$ the occupation of the corresponding virtual queue. The resulting policy $\pi_V^\star$ chooses the action as follows:
$$\pi_V^\star (s) = \min \left\{ \min_{i \in \{1, \dots, M \}}  \left \{ \frac{\tilde{s}_i}{\mu_i'} \right\}, \min_{i \in \{ M+1, \dots, N \}} \left \{ \frac{s_i}{\mu_i} \right\} \right\}$$
The policy $\pi_V^\star$ is defined as JSVED (Join the Shortest Virtual Expected Delay). By construction, it is immediate to show that if $\pi_V^\star$ renders $\tM$ stable, it also ensures constrains are respected in $\M$. 

The main limitation of JSVED is that it tends to be conservative. In fact, whenever $\theta_i \leq 1/\mu_i$, we have $\tilde{s}_i \leq \tilde{s}_i$ w.p.1. Because JSVED determines its actions based on the occupation levels of the virtual queues, the chosen action may be optimal with respect to the virtual queues, but suboptimal concerning the actual state of the real queues. The two safe policies we introduce in the next section guarantee safety, but rely on the real occupation vector only.
%%%%%%%%%%%%%%%%%%%%%%%%%%%%%%%%%%%%%%%%%%%%%%%%%%%%%%%%%%%%
\subsection{JSED-$k$: JSED with memory}
\label{subsec:JSEDk}
%%%%%%%%%%%%%%%%%%%%%%%%%%%%%%%%%%%%%%%%%%%%%%%%%%%%%%%%%%%%
The second policy we propose is similar to the approach used in JSED, but with a key difference: it stores the last $k$ actions in memory. This allows the system to estimate the average cost at the time when an action is taken. Based on this estimate, the action space for the current timestep is constructed to prioritize safety. More precisely, at each step, the policy selects a queue that has been chosen no more than $k \theta_i$ times within the last $k$ steps. 

JSED-$k$ evaluates the average constraint in the last $k$-step window: for sufficiently large $k$, the policy is safe, as indicated by the following result. The full proof, just sketched here, can be found 
\ifextver
in Appendix \ref{proof:JSEDk}.
\else
in \cite{extendedVersion}.
\fi
\begin{proposition}
The policy JSED-$k$ is safe for a sufficiently high value of $k$. 
\label{prop:JSEDk safety}
\end{proposition}
\begin{IEEEproof} First, observe that $k$ determines the accuracy of the average cost estimate, which in turn defines the action space at each timestep. Moreover, for $k = \infty $, i.e., when the entire past history is used to define the action space at a given timestep, the policy is seen to be safe. Hence, for all $\epsilon > 0 $, there exists a value of $k$ such that the constraint is violated with probability at most $\epsilon$.   
\end{IEEEproof}
The JSED-$k$ approach may require significant memory resources to store a sufficiently long action history. As the number of queues in the system increases, the required memory size $k$ must grow at least linearly to provide an accurate estimate of the average cost. In \cref{sec:numerical results}, we study numerically the relationship between the number of queues $ N $ and the memory size $k$ required to ensure safety.
%%%%%%%%%%%%%%%%%%%%%%%%%%%%%%%%%%%%%%%%%%%%%%%%%%%%%%%%%%%%
\subsection{JSSQ: Join the Shortest Safe Queue}
\label{subsec:JSSQ}
%%%%%%%%%%%%%%%%%%%%%%%%%%%%%%%%%%%%%%%%%%%%%%%%%%%%%%%%%%%%
The last policy does not need to store the last $k$ actions in memory and yet makes decisions directly based on the actual occupancy of the queues. This dynamic policy sets a target average arrival rate per queue to respect the average constraints and minimize the average occupation. Actions are decided at each timestep based on the current server occupation and the corresponding target value $L_i$.

To find the desired arrival rate for each queue, first consider the nonlinear problem describing the system behavior, whose objective is to minimize the average system occupation:
\begin{align}
\label{eq:system for minimizing objective function}%\tag{TR}
    &\underset{\xi}{\text{minimize: }} && \sum_i \frac{\xi_i}{\mu_i - \xi_i} \\
    &\text{subj. to: } && \xi_i < \mu_i ' \quad \forall i, \notag \\
    &&& \sum_i \xi_i = \Xi \notag,
\end{align}
where $\Xi$ represents the total arrival rate in the system. In the stability region, \ref{eq:system for minimizing objective function} is a convex optimization problem as the objective function $\frac{\xi_i}{\mu_i - \xi_i}$ is convex for $\xi_i < \mu_i$. Standard convex optimization methods %, such as gradient descent or Lagrange multipliers, 
can be employed to solve this problem efficiently. Once the optimal arrival rates $\xi_i$ are determined, the corresponding target queue lengths $L_i$ can be computed as $L_i := \frac{\xi_i}{\mu_i - \xi_i}$
where $\xi_i$ is the solution of \eqref{eq:system for minimizing objective function} for queue $i$, and $\mu_i$ is its processing rate.  

Once the desired values of $L_i$ are known, we can define a policy whose primary goal is to maintain the occupancy of each queue as close as possible to its optimal level while satisfying the constraints. 

%By achieving this balance, the policy prevents queues from becoming overloaded or underutilized, ensuring system stability and smooth operation. This approach guarantees safety by adhering to predefined constraints while achieving efficient queue management, reducing delays, and improving throughput.
At each timestep, the policy operates as follows. If at least one queue has an occupancy below its desired average, the policy selects a queue from this subset $G$. The selection is made with probability  proportional to the desired arrival rates, $\xi_i/\sum_{j \in G} \xi_j$. In the case where every queue $i$ exceeds their target average occupations, i.e., $s_i \geq L_i$, the policy employs a randomized assignment dispatching an arrival to queue $i$ with probability $\xi_i/\Xi$.%the queues according to the proportion implied by their corresponding values of $\xi_i$. %The approach balances queue occupancies while adhering to safety constraints.  

The safety of this policy can be established. The proof, just sketched hereafter, is detailed  
\ifextver
in Appendix \ref{proof:JSSQ}.
\else
in \cite{extendedVersion}.
\fi  
\begin{proposition}  
JSSQ is safe.  
\label{prop:JSSQ safety}
\end{proposition}  
\begin{IEEEproof}  
If a queue's occupancy is below its desired average, it implies that the arrival rate to that queue has been lower than its target rate and, consequently, lower than the constraint associated with that queue. %Adding arrivals to one of these queues will not violate the safety requirements. 
Conversely, if all queues have occupancies above their desired averages, the randomization step is performed according to a safe policy.  
\end{IEEEproof}  
This policy presents several advantages compared to the methods discussed before. First, it enables real-time decision-making by acting on the actual state of the system, unlike JSVED. Second, it achieves memory efficiency, since it does not require storing the last $k$ actions, making it computationally more efficient than JSED-$k$. This policy can also be easily adapted, with minimal modifications, to solve the constrained version of the unconstrained problem analyzed in \cite{zhouAsymptoticallyOptimalLoad2021,zhouHeavytrafficDelayOptimality2018a,zhouDegreeQueueImbalance2018,zhouDesigningLowComplexityHeavyTraffic2017}. This suggests a potential application of JSSQ in the context of those works, where the focus is the scalability to a large number of queues  

JSSQ adresses the constraint satisfaction using randomization. As the other proposed heuristics, this approach does not guarantee the optimality of the policy, which, as remarked before, requires rather to optimize the Lagrange multipliers. Indeed, a promising avenue for future work is to refine JSSQ by integrating approximations of the Lagrange multipliers described in \cref{sec:optimal policy properties}. This enhancement could result in a policy that closely approximates the optimal solution while maintaining the safety guarantees, and it is left as future work.  
%%%%%%%%%%%%%%%%%%%%%%%%%%%%%%%%%%%%%%%%%%%%%%%%%%%%%%%%%%%%%%%%%%%%%%%%%%%%%%%%%%%%%%%%%%%%%%%%%%%%%%%%%%%%%%%%%%%%%%%%
%%%%%%%%%%%%%%%%%%%%%%%%%%%%%%%%%%%%%%%%%%%%%%%%%%%%%%%%%%%%%%%%%%%%%%%%%%%%%%%
\section{Ensuring Resource Efficiency Through Minimum Occupation Constraints}
\label{sec:generic cost constraints}
%%%%%%%%%%%%%%%%%%%%%%%%%%%%%%%%%%%%%%%%%%%%%%%%%%%%%%%%%%%%%%%%%%%%%%%%%%%%%%%
%%%%%%%%%%%%%%%%%%%%%%%%%%%%%%%%%%%%%%%%%%%%%%%%%%%%%%%%%%%%%%%%%%%%%%%%%%%%%%%%%%%%%%%%%%%%%%%%%%%%%%%%%%%%%%%%%%%%%%%%
In this section, we consider a different system in which the gain function depends on the stqte of the system. Specifically, we assume that each queue $i$ must satisfy an average constraint of the form:
\begin{equation}
    K_\pi^i(\beta) = \expvalDist{\pi, s_0 \sim \beta}{\lim_{T \to \infty} \frac{1}{T}\sum_{t=0}^T c_i(S_t) \mid S_0 = s_0} \geq \theta_i
    \label{eq:generic constraint}
\end{equation}
for a generic convex gain function $c_i(\cdot)$.
% \textbf{Quality of service constraint:} it ensures that system resources are utilized efficiently to meet user expectations and performance standards, preventing under-provisioning that can cause delays or failures, and over-provisioning that wastes resources and increases costs.

% Both these constraints and the one previously considered work to balance system performance. QoS constraints aim to ensure satisfactory user experience, while link capacity constraints prevent overloading of system components, which could degrade QoS.

In the remainder of this section, we will analyze how JSED-$k$ and JSSQ, two of the policies introduced in \cref{sec:safe policies} to solve the initial problem, can be applied to the one considering the constraint in \eqref{eq:generic constraint}. We will exclude JSVED from the discussion due to its poor performance (see Section \ref{sec:numerical results}).

JSED-$k$ extends naturally to the general case by approximating the gain function in a manner similar to how it is handled for constraints of type \eqref{eq:immediate cost}. % However, the structure of the value function may impact significantly the relationship between the memory required to ensure safety and the system size.

Finally, we will demonstrate how to construct the JSSQ policy for the problem with a generic constraint of the type \eqref{eq:generic constraint}. As outlined in \cref{subsec:JSSQ}, the first step is to determine a value for the target arrival rate $\xi_i$ that satisfies a nonlinear problem characterizing the case under study: 
\begin{align}
\label{eq:system for minimizing generic function}
    &\underset{\xi}{\text{minimize: }} && \sum_i \frac{\xi_i}{\mu_i - \xi_i} \\
    &\text{subject to} && c_i\rp{\frac{\xi_i}{\mu_i - \xi_i}} > \theta_i \quad \forall i, \notag \\
    &&& \xi_i < \mu_i \quad \forall i, \notag \\
    &&& \sum_i \xi_i = \Xi \notag,
\end{align}
where $\theta_i$ is now the generic constraint and $c_i \rp{\frac{\xi_i}{\mu_i - \xi_i}}$ is the gain function associated to the $i$-th queue computed in the average state.  
Clearly, the constraints related to flow conservation and those related to stability still need to be verified.% to ensure comprehensive system performance and reliability.

Once the desired values of $\xi_i$ are known, the policy is defined as in \cref{subsec:JSSQ}, i.e., JSSQ tries to keep the occupation of the MDP as close as possible to the one induced by the solution of \eqref{eq:system for minimizing generic function}.

The safety of this policy is an immediate consequence of Jensen inequality:
\begin{corollary}
    If a policy induces a set of arrival rates that satisfy \eqref{eq:system for minimizing generic function}, then the policy is also safe.
\end{corollary}
\begin{IEEEproof}
    In \eqref{eq:system for minimizing generic function} the first condition to satisfy involves the gain of the average state obtained by a policy, while in the MDP we have to consider the average gain of a policy. By Jensen inequality  $c\rp{\expvalDist{}{(s)}} \leq \expvalDist{}{c(s)}$ as the gain function $c(\cdot)$ is assumed to be convex. Therefore, if we can guarantee that  $c_i\rp{\expvalDist{}{s}} > \theta_i \  \forall i$, we know that the constraint is satisfied if the queue has arrival rate $\xi_i$. 
\end{IEEEproof}

%%%%%%%%%%%%%%%%%%%%%%%%%%%%%%%%%%%%%%%%%%%%%%%%%%%%%%%%%%%%%%%%%%%%%%%%%%%%%%%%%%%%%%%%%%%%%%%%%%%%%%%%%%%%%%%%
%%%%%%%%%%%%%%%%%%%%%%%%%%%%%%%%%%%%%%%%%%%%%%%%%%%%%%%%%%%%%%%%%%%%%%%%%%%%%%%
\section{Numerical results}
\label{sec:numerical results}
%%%%%%%%%%%%%%%%%%%%%%%%%%%%%%%%%%%%%%%%%%%%%%%%%%%%%%%%%%%%%%%%%%%%%%%%%%%%%%%
%%%%%%%%%%%%%%%%%%%%%%%%%%%%%%%%%%%%%%%%%%%%%%%%%%%%%%%%%%%%%%%%%%%%%%%%%%%%%%%%%%%%%%%%%%%%%%%%%%%%%%%%%%%%%%%%
Numerical results are divided into four main groups. 

The first one explores the computational complexity of the Lagrange multipliers optimization, i.e., the additional step required to obtain a near-optimal policy described in \cref{sec:optimal policy properties}.
In the second set of experiments, we study the performance of JSED-$k$ for increasing $k$. We specifically focus on the memory size required for the resulting policy to be safe. The third experiment compares the performance of the policies presented in \cref{sec:safe policies} - namely, JSVED, JSED-$k$, and JSSQ - with JSED, the state-of-the-art algorithm conceived for unconstrained load balancing.
The last experiment provides a comparative analysis of the algorithms discussed in \cref{sec:numerical results}, dealing with a cost function which  depends on the current state.

We randomly sampled the system parameters from predefined sets for all experiments and normalized  their outcomes accordingly.%Each column in the table represents the sets considered to generate the corresponding data.

Throughout this section, we considered either light or heavy load conditions to represent different system load conditions. Under light load conditions, the arrival rate of tasks or requests falls between $70\%$ and $95\%$ of the total processing capacity of the system. These conditions represent a system operating with sufficient spare capacity, experiencing manageable workloads and relatively small delays. This is meant to measure the system performance under relatively relaxed operating conditions, providing insight into efficiency and responsiveness when resources are not critically constrained.

Under heavy load conditions, the arrival rate exceeds $99\%$ of the system's total processing capacity. This reflects near-saturation levels, where the system is pushed to its operational limits. Heavy load conditions are particularly relevant for stress-testing the system's ability to maintain performance under high demand. %, highlighting potential bottlenecks, delays, or failures. 
These scenarios are critical to understand the robustness and reliability of the system in real-world situations where demand may approach or even exceed design thresholds.
%By analyzing both light and heavy traffic conditions, we aim to provide a comprehensive understanding of the system's behavior across a broad spectrum of operational contexts, offering valuable insights into its scalability, efficiency, and resilience.
%%%%%%%%%%%%%%%%%%%%%%%%%%%%%%%%%%%%%%%%%%%%%%%%%%%%%%%%%%%%%%%%%%%%%%%%%%%%%%%
\subsection{Optimizing Lagrange multipliers}
%%%%%%%%%%%%%%%%%%%%%%%%%%%%%%%%%%%%%%%%%%%%%%%%%%%%%%%%%%%%%%%%%%%%%%%%%%%%%%%
The first experiment reports on the number of policy evaluation steps needed to determine the optimal Lagrange multipliers  as described in \eqref{eq:greedy action Lyapunov}. We considered environments with light traffic conditions and $5$, $10$, and $25$ queues. The results shown in \cref{tab:LM optimization numerical results} are obtained from $30$ independent experiments per value of $N$. We can observe that, even for smaller values of $N$, the convergence speed, measured by the required policy evaluation steps, remains remarkably high. We set an upper limit of $50$ for the number of policy evaluations. It is worth observing that, in these tests, the number of policy evaluation steps using unsafe policies was very close to the total. This finding highlights a critical limitation of this approach, particularly in scenarios where evaluating unsafe policies online could lead to undesirable consequences. 
\begin{table}
    \centering
    \begin{tabular}{|c|c|c|c|}
    \hline
        \rowcolor{gray!30}$\boldsymbol{N}$ & $\boldsymbol{5}$ & $\boldsymbol{10}$ & $\boldsymbol{25}$ \\
        \hline
        %\textbf{Mean convergence speed} & 16.475 & 24.9 & 41.72 \\
        \textbf{mean \# of iterations} & 16.475 & 24.9 & 41.72 \\
        \hline
        %\textbf{\# of occurrences with $\boldsymbol{>50}$ iterations }& 5 & 9 & 18\\
        \textbf{occurrences with $\boldsymbol{>50}$ iterations }& $17\%$ & $30\%$ & $60\%$\\
        \hline
    \end{tabular}
    \caption{Number of iterations for the stochastic approximation method optimizing the Lagrange multipliers; the policy chooses actions according to \eqref{eq:greedy action Lyapunov}; low traffic conditions.}
    \label{tab:LM optimization numerical results}
\end{table}

%%%%%%%%%%%%%%%%%%%%%%%%%%%%%%%%%%%%%%%%%%%%%%%%%%%%%%%%%%%%%%%%%%%%%%%%%%%%%%%
\subsection{Performance of JSED-$k$: different memory size}
\label{subsec:numerical JSEDk}
%%%%%%%%%%%%%%%%%%%%%%%%%%%%%%%%%%%%%%%%%%%%%%%%%%%%%%%%%%%%%%%%%%%%%%%%%%%%%%%
\begin{figure}[t]
    \centering
    \includegraphics[width=0.85\linewidth]{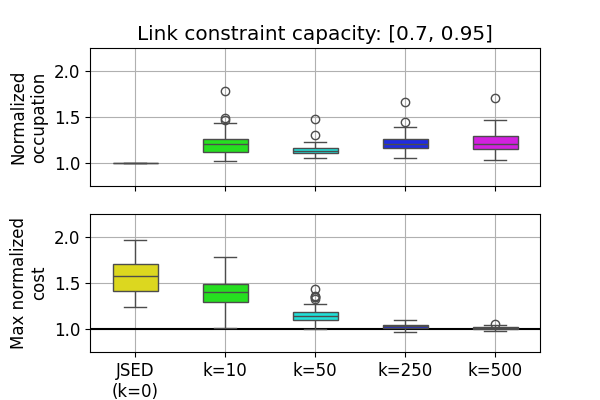}
    \caption{Average occupation and max cost of JSED-$k$ for increasing values of memory size $k$; light traffic conditions.}
    \label{fig:comparison memory size}
\end{figure}

\begin{figure}
    \centering
    \includegraphics[width=0.85\linewidth]{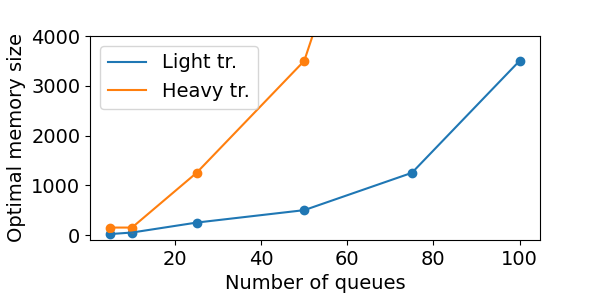}
    \caption{Evolution of the critical memory size at the increase of the number of queues.} %Both light and heavy traffic conditions are represented.}
    \label{fig:optimal safe memory size}
\end{figure}
In this section, we analyze JSED-$k$ for increasing memory size $k$. A larger memory size reduces the probability of violating the constraints. But, it introduces practical challenges, as discussed in \cref{subsec:JSEDk}.  

We reported our findings in \cref{fig:comparison memory size} and \cref{fig:optimal safe memory size}. The first one compares the reward and the maximum cost for increasing memory size. The system has $N=25$ queues, $M=13$ of which are constrained. The second figure shows the evolution of the critical memory size, i.e., the smallest memory size $k$ that guarantees safety as as function of the number of queues.

\Cref{fig:comparison memory size} presents data from a system operating under light traffic conditions, i.e., the arrival rate does not exceed $95\%$ of the system's total capacity. As expected, increasing $k$ reduces the cost associated with the queue experiencing the highest cost, bringing the system closer to satisfying all constraints. Notably, for $k =250$, even the highest observed cost across all experiments attains, on average, the constraint threshold.  

In \cref{fig:optimal safe memory size}, we analyze how the critical memory size evolves as the number of queues in the system increases, with $M=N/2$ %the number of constrained queues set to half of the total 
for simplicity. The results indicate that in both load conditions, the critical memory size follows an exponential growth pattern as $N$ increases. However, for light load conditions, the critical memory size remains practically viable, namely up to $100$ queues. %even for systems with a large number of queues. 
Conversely, for heavy load conditions, the critical memory constraints grows significantly faster, becoming impractical even for relatively small systems. Larger values of $N$ ($N>50$) required $k$ to exceed $10,000$, which was also the maximum value we tested. We excluded those data points for the sake of readability.

%%%%%%%%%%%%%%%%%%%%%%%%%%%%%%%%%%%%%%%%%%%%%%%%%%%%%%%%%%%%%%%%%%%%%%%%%%%%%%%
\subsection{Comparison performances of different safe policies}
%%%%%%%%%%%%%%%%%%%%%%%%%%%%%%%%%%%%%%%%%%%%%%%%%%%%%%%%%%%%%%%%%%%%%%%%%%%%%%%
\begin{figure}
    \centering
    \includegraphics[width=0.95\linewidth]{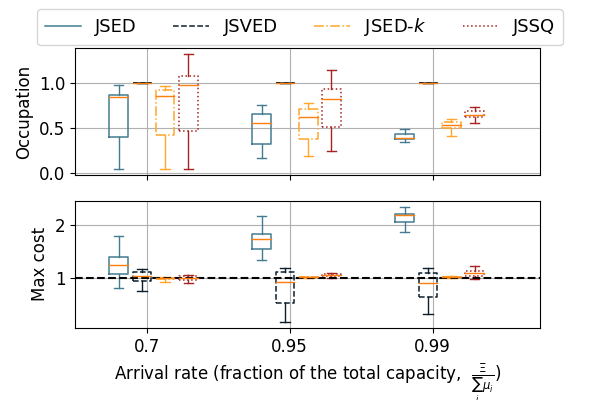}
    \caption{Comparison of the policies described in \cref{sec:safe policies} for increasing ratio between arrival rate and capacity of the system.}% The results are averaged over 50 experiments, each conducted with independently drawn parameters.}
    \label{fig:comparison safe policies}
\end{figure}

We then analyzed the performance of the methods introduced in \cref{sec:safe policies}, comparing them to JSED, the state-of-the-art algorithm for unconstrained load balancing. The comparison is carried out under both heavy and light traffic conditions by varying the ratio between the arrival rate and the system's cumulative processing rate. We fixed the same set of $50$ environments to test all policies; the arrival rate, on the contrary, was progressively increased.

For each method, \cref{fig:comparison safe policies} presents the distribution of the normalized occupation  and the normalized maximum cost. The occupation has been normalized with respect to the value obtained by JSVED. Cost normalization is performed component by component relative to the value of the constraint. A method is considered safe if the cost associated with each queue remains below its corresponding constraint. In \cref{fig:comparison safe policies}, we emphasize the value of the worst-performing queue observed in each experiment to illustrate the safety of the policies.
%A method is deemed safe if the cost associated with the worst-performing queue is bounded below the constraint threshold.

As expected, JSED achieves the lowest average system occupation across all experiments, as it is provably the optimal policy in the unconstrained case. However, it fails to respect the constraints, particularly in scenarios with heavier traffic. In contrast, the results clearly show that the safe policies introduced in \cref{sec:safe policies} successfully satisfy the constraints across a range of arrival rate values.

Regarding average system occupation, JSVED performs the worst, as discussed in \cref{subsec:JSVED}, with its limitations becoming even more pronounced under heavier traffic conditions. Finally, JSED-$k$, when equipped with an appropriately sized memory as outlined in \cref{subsec:numerical JSEDk}, demonstrates superior resource management compared to JSSQ, striking a balance between performance and constraint compliance.

By combining the results discussed in this section and those in \cref{subsec:numerical JSEDk}, we conclude that for relatively small systems, JSED-$k$ emerges as the most effective safe policy in terms of minimizing average system occupation, with JSSQ showing slightly lower performance.  We recall, however, that even under heavy traffic conditions, JSSQ has $O(1)$ memory requirements. Hence, JSSQ is preferable once memory allocation or computation of actions' statistics required by JSED-$k$ become impractical.  
%The results clearly demonstrate that the safe policies introduced in \cref{sec:safe policies} satisfy the constraints for a range of arrival rate values. These numerical experiments suggest that, for relatively small systems, JSED-$k$ is the most effective safe policy in terms of average occupation, with JSSQ showing slightly lower performance. We recall, however, that even under heavy traffic conditions, JSSQ has $O(1)$ memory requirements. Hence, it may be preferable when memory allocation or the computation of actions' statistics required by JSED-$k$ become impractical.  
 
%%%%%%%%%%%%%%%%%%%%%%%%%%%%%%%%%%%%%%%%%%%%%%%%%%%%%%%%%%%%%%%%%%%%%%%%%%%%%%%
%%%%%%%%%%%%%%%%%%%%%%%%%%%%%%%%%%%%%%%%%%%%%%%%%%%%%%%%%%%%%%%%%%%%%%%%%%%%%%%
\subsection{State-dependent cost function}
%%%%%%%%%%%%%%%%%%%%%%%%%%%%%%%%%%%%%%%%%%%%%%%%%%%%%%%%%%%%%%%%%%%%%%%%%%%%%%%
\begin{figure}
    \centering
    \includegraphics[width=0.85\linewidth]{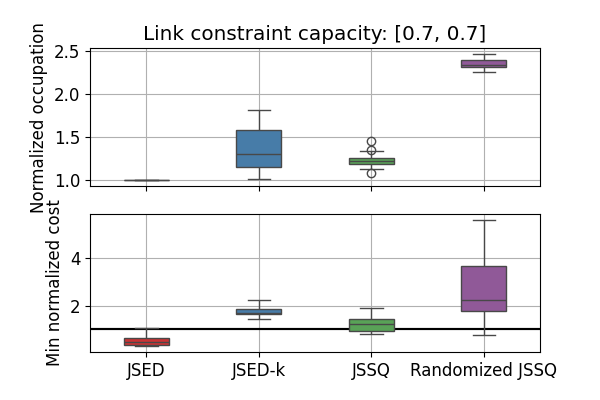}
    \caption{Comparison of the reward and safety guarantees for different policies in the case of a convex generic gain function.}
    \label{fig:exp 3 convex cost function}
\end{figure}

\begin{comment}
    \centering
    \includegraphics[width=0.85\linewidth]{numerical_results/generic_linear_cost_comparison_methods_light_traffic.png}
    \caption{Linear cost function}
    \label{fig:exp 3 linear cost function}
\end{comment}

The numerical section concludes by analyzing the performance of the methods introduced in \cref{sec:generic cost constraints}. Additionally, we evaluated the performance of a randomized policy that selects a server based on probabilities induced by the arrival rates obtained by solving \eqref{eq:system for minimizing generic function}. This policy is referred to as \textit{Randomized JSSQ}. As expected, it performed consistently worse than JSSQ across all experiments.

The experiments were conducted with $N = 10$ servers, for linear and convex gain functions. % , where the argument of the gain function is the occupation of each queue. 
The same set of $50$ environments was used in both cases. %for both gain function types, differing only in the definition of the gain function. 
For sake of space, and because of the minor differences between the two cases, only the results for the convex gain function are presented here. All experiments were performed under light traffic conditions.

In \cref{fig:exp 3 convex cost function}, we compare the normalized occupation and the minimum normalized gain across all queues. A policy is considered safe if the minimum gain exceeds the unitary normalized constraint (all values are normalized).

In terms of occupation, JSED performs the best, as expected, while JSSQ slightly outperforms JSED-$k$. The  policies, as defined in the experiments, remain safe throughout all tests. JSSQ consistently achieves performance close to JSED in terms of occupation, while JSED-$k$ exhibits higher variance. JSED-$k$'s performance is more sensitive to changes in the environment, unlike JSSQ, which shows greater stability.

%%%%%%%%%%%%%%%%%%%%%%%%%%%%%%%%%%%%%%%%%%%%%%%%%%%%%%%%%%%%%%%%%%%%%%%%%%%%%%%%%%%%%%%%%%%%%%%%%%%%%%%%%%%%%%%%%%%%%%%%
%%%%%%%%%%%%%%%%%%%%%%%%%%%%%%%%%%%%%%%%%%%%%%%%%%%%%%%%%%%%%%%%%%%%%%%%%%%%%%%
\section{Related works}
\label{sec:related works}
%%%%%%%%%%%%%%%%%%%%%%%%%%%%%%%%%%%%%%%%%%%%%%%%%%%%%%%%%%%%%%%%%%%%%%%%%%%%%%%
%%%%%%%%%%%%%%%%%%%%%%%%%%%%%%%%%%%%%%%%%%%%%%%%%%%%%%%%%%%%%%%%%%%%%%%%%%%%%%%%%%%%%%%%%%%%%%%%%%%%%%%%%%%%%%%%%%%%%%%%

State-dependent routing problems have been extensively studied in the literature (see, e.g., \cite{argon2009dynamic} for an overview). A notable setting where simple structures for optimal policies have been successfully identified involves homogeneous stations. In such cases, the ``Shortest Delay Routing'' (SDR) policy has been demonstrated to be optimal under various conditions \cite{winston1977optimality, weber1978optimal}.
However, \cite{whitt1986deciding} showed also several trivial cases in which this policy is not optimal.
For the more complex case of routing problems that do not assume station homogeneity, \cite{foschini1977heavy} showed the asymptotical optimality of the JSED policy in the heavy traffic case. When the cost function is a convex function of occupation, \cite{mandelbaum2004scheduling} proves the optimality under heavy traffic of a policy which chooses at each step the greedy action minimizing $c_i(s_i) \mu_i$, i.e., the immediate cost multiplied by the processing speed of the queue.
\cite{krishnanJoiningRightQueue1987} defines a near-optimum queue-assignment rule that offers a generalization of the shortest-queue rule to the case of dissimilar queues to minimize the average sojourn time. \cite{buyukkoc1985cmu} provides an optimal policy when $N$ queues compete for a single server (as in call centers). Finally, \cite{stolyar2005optimal} proposes an optimal solution in the case of non-linear cost functions. 
\begin{comment}
\paragraph{Queues with admission cost (pricing)}
\cite{tandra2004join}

\cite{dubeDifferentialJoinPrices2002}

\cite{bodasLoadBalancingRouting2011}

\cite{bradfordPricingRoutingIncentive1996}
    
\end{comment}
%The Lyapunov drift-plus-penalty optimization is a popular technique that can be applied to queueing networks and other stochastic systems and has been introduced in \cite{neely2010queue}. Due to its simplicity and efficiency it has been applied in many scenarios, such as routing, IoT and adaptive video streaming. 
To the best of the authors' knowledge, despite being a problem of clear practical interest, the constrained version of the load balancing problem has not been tackled so far in the literature. 

%%%%%%%%%%%%%%%%%%%%%%%%%%%%%%%%%%%%%%%%%%%%%%%%%%%%%%%%%%%%%%%%%%%%%%%%%%%%%%%%%%%%%%%%%%%%%%%%%%%%%%%%%%%%%%%%%%%%%%%%
%%%%%%%%%%%%%%%%%%%%%%%%%%%%%%%%%%%%%%%%%%%%%%%%%%%%%%%%%%%%%%%%%%%%%%%%%%%%%%%
\section{Conclusions}
%%%%%%%%%%%%%%%%%%%%%%%%%%%%%%%%%%%%%%%%%%%%%%%%%%%%%%%%%%%%%%%%%%%%%%%%%%%%%%%
%%%%%%%%%%%%%%%%%%%%%%%%%%%%%%%%%%%%%%%%%%%%%%%%%%%%%%%%%%%%%%%%%%%%%%%%%%%%%%%%%%%%%%%%%%%%%%%%%%%%%%%%%%%%%%%%%%%%%%%%
Load balancing is critical in many applications to ensure efficient resource allocation and system performance. However, standard load-balancing algorithms are oblivious to constraints such as link capacities or state-dependent restrictions on individual queues. As a result, while similar constraints are the norm in real-world systems, current load-balacing methods are not designed to address them.

%In this work, we proposed new policies to perform %the load balancing problem while incorporating these realistic, complex constraints. 
%load balancing and incorporate constraints at once. We first identified a near-optimal method based on CMDP relaxation. While the CMDP framework provides fine theoretical insight, it is hardly viable for real-world scenarios since it entails policy evaluation steps which are computationally demanding. 
In this paper, we first identified a near-optimal method based on a CMDP relaxation. While it provides theoretical insight, the method is hardly viable for real-world scenarios since it entails policy evaluation steps which are computationally demanding. Hence, we proposed new lightweight policies to perform load balancing and incorporate constraints at once.

This work is just an initial step in constrained load balancing. Future research will focus on simplifying the near-optimal approach using fast approximation methods to estimate optimal Lagrange multipliers. %Additionally, in heavy traffic, scalability may depend on message complexity and on delay constraints, as the occupancy of the queues needs to be sampled in real time. 
Also, in heavy traffic, scalability depends on message complexity and delay constraints, as queue occupancy must be sampled in real time. Solving these open challenges will broaden the applicability of constrained load balancing solutions to real-world scenarios.
\bibliographystyle{ieeetr}
\bibliography{biblio.bib}

\ifextver
\newpage
\onecolumn
\appendix 
%%%%%%%%%%%%%%%%%%%%%%%%%%%%%%%%%%%%%%%%%%%%%%%%%%%%%%%%%%%%%%%%%%%%%%%%%%%%%%%
\subsection{Proof of \cref{prop:Lyapunov method}}\label{proof:lyap}
%%%%%%%%%%%%%%%%%%%%%%%%%%%%%%%%%%%%%%%%%%%%%%%%%%%%%%%%%%%%%%%%%%%%%%%%%%%%%%%

\begin{IEEEproof} We write the Lyapunov function of the system \cite{neely2022stochastic} as 
\begin{align}
    L(s(t)) = \frac{1}{2} \sum_{i=1}^N \frac{s_i(t)^2}{\mu_i}
\end{align}
and the corresponding conditional Lyapunov drift in slot $t$ writes 
\begin{equation}
\Delta (s(t)) = \expvalDist{}{L(s(t+1)) - L(s(t)) \mid s(t)}
\end{equation}
The proof of \cref{prop:Lyapunov method} is an immediate consequence of the fundamental result in \cite{neely2022stochastic}[Thm. 4.2] that we recall here for the sake of reading: 
%The following result is verified:
\begin{theorem}
    If $\epsilon \geq 0, V\geq 0, B\geq 0$ exist such that for all $t$ slots we have 
\begin{equation}
    \Delta(s(t)) + V \expvalDist{}{\,r(t) \mid s(t)} \leq B + V r^\star - \epsilon \sum_{i=1}^N \frac{s_i(t)}{\mu_i}
    \label{eq:bound optimization}
\end{equation}
then all queues $s_i(t)$ are (mean-rate) stable and the time-averaged expected penalty $\overline{r}$ are bounded by 
\begin{align}
    \overline{r} &\leq r^\star + \frac{B}{V}\\
    \lim_{T \rightarrow \infty} \frac{1}{T} \sum_{\tau = 0}^{N-1} \sum_{i=1}^N s_i(\tau) &\leq \frac{B + V(r^\star - r_{min})}{\epsilon}
\end{align}
where $B$ is a positive constant that upper bounds the expression $\expvalDist{}{\frac{\alpha(t)^2 + \eta(t)^2}{2} \mid s(t)}$
\label{thm:Lyapunov optimization}
\end{theorem}

It remains to prove that we can apply \cref{thm:Lyapunov optimization} and in particular that \eqref{eq:bound optimization} is verified.

We observe how 
\begin{align*}
\Delta(x(t)) &= \expvalDist{}{L(s(t+1)) - L(s(t))} = \expvalDist{}{\frac{1}{2} \sum_{i=1}^N \frac{s_i(t+1)^2}{\mu_i} - \frac{s_i(t)^2}{\mu_i}}\\
&=  \sum_i \frac{1}{2 \mu_i} \expvalDist{}{ \alpha_i^2 (t) + \eta_i^2(t) + 2 s_i(t) \alpha_i(t) - 2 s_i(t) \eta_i(t) - 2 \alpha_i(t) \eta_i(t)}\\
&\leq \sum_i  \frac{1}{2} \expvalDist{}{\alpha_i^2(t) + \eta_i^2(t)} + \sum_i \frac{1}{2 \mu_i} \expvalDist{}{2 s_i(t) (\alpha_i(t) - \eta_i(t))}\\
&\leq B + \sum_i \frac{1}{2 \mu_i} \expvalDist{}{2 s_i(t) (\alpha_i(t) - \eta_i(t))}\\
&\leq B - \epsilon \sum_i \frac{s_i}{\mu_i} 
\end{align*}
with the last inequality following from the fact that the expected $\expvalDist{}{\alpha_i(t) \mid s_i(t)} \leq \expvalDist{}{\eta_i(t) \mid s_i(t)}$ is a necessary condition for the stability of the system.\end{IEEEproof}
%%%%%%%%%%%%%%%%%%%%%%%%%%%%%%%%%%%%%%%%%%%%%%%%%%%%%%%%%%%%%%%%%%%%%%%%%%%%%%%
\subsection{Proof of \cref{prop:JSEDk safety}}\label{proof:JSEDk}%\label{proof:lyap}
\begin{IEEEproof}
%%%%%%%%%%%%%%%%%%%%%%%%%%%%%%%%%%%%%%%%%%%%%%%%%%%%%%%%%%%%%%%%%%%%%%%%%%%%%%%
We want to prove that for every $\epsilon > 0$ there exists $k$ such that
\[
J_{\pi_k, i}(\beta) \leq \theta_i + \epsilon 
\]
where $\pi_k$ denotes the JSED-$k$ policy, assuming memory size $k$, and $J_{\pi_k, i}$ is the corresponding reward.

First, we observe that the policy $\tilde{\pi}$, which considers the entire past history to determine the action set, is safe. In particular, at each timestep, it can determine which actions would not violate the constraint if chosen.

Then, we observe that, given a certain state $s_t$ and the past actions, the estimator $\tilde{c}_{k, i} (s_t) = \sum_{j=1}^k \frac{\delta_i(a_{t-k+1})}{k}$ is a consistent estimator of the average cost for the full trajectory, denoted as %
$c_i (s_t)$. This implies that, for every $\Delta>0$, $$\lim_{k\to \infty} \Pro{\mid c_{k, i} - c_i  \mid < \Delta } = 1$$
This further implies that, for every fixed $\Delta$ there exists $k \in \mathbb{N}$ such that for every $\epsilon>0$ it holds  
\begin{equation}
    %\forall \epsilon, \ \exists k : \Pro{\mid c_{k, i} - c_i \mid > \epsilon'} < \epsilon 
    \Pro{\mid c_{k, i} - c_i \mid > \Delta} < \epsilon 
    \label{eq:consistent estimator}
\end{equation}

Now, fixed $\epsilon> 0$ and $k=k(\Delta) \in \mathbb{N}$ defined above it holds
\begin{align}
    \left | J_{\pi_k, i} (\beta) - J_{\tilde{\pi}, i}(\beta) \right | &= \left |\expvalDist{a_t \sim \pi_k, s_0 \sim \beta}{\lim_{T \to \infty} \frac{1}{T} \sum_{t=1}^T c_i (s_t, a_t) } - \expvalDist{a_t \sim \tilde{\pi}, s_0 \sim \beta}{\lim_{T \to \infty} \frac{1}{T} \sum_{t=1}^T c_i (s_t, a_t) }\right |\notag \\
    & = \left | \expvalDist{a_t \sim \pi_k, s_0 \sim \beta}{\lim_{T \to \infty} \frac{1}{T} \sum_{t=1}^T \delta_i(a_t) } - \expvalDist{a_t \sim \tilde{\pi}, s_0 \sim \beta}{\lim_{T \to \infty} \frac{1}{T} \sum_{t=1}^T \delta_i(a_t) } \right |\notag \\
    & \leq \expvalDist{s_0 \sim \beta}{\lim_{T \to \infty} \frac{1}{T} \sum_{t=1}^T \mid \rp{1 -\delta_{\pi_k(s_t)}(\tilde{\pi}(s_t) }  \cdot \rp{ \delta_i \rp{\pi_k(s_t)}  + \delta_i(\tilde{\pi}(s_t)}  \mid } \label{eq:proof prop 2 inequality}\\
    & \leq \expvalDist{}{\lim_{T\to\infty} \frac{1}{T} \sum_{t=1}^T \boldsymbol{1} \rp{\mid c_{k, i}(s_t) - c_i(s_t) \mid > \Delta } } \notag \\
    & = \Pro{\mid c_{k, i}(s_t) - c_i(s_t) \mid > \Delta }\notag
\end{align}
%where the delta function is defined as the function being one when the two arguments are equal and $0$ otherwise while 
where each term appearing in the sum in \eqref{eq:proof prop 2 inequality} %is %a function 
is equal to one if and only if a policy chooses action $i$ and the other does not,  and it is zero otherwise.

% Then, due to \eqref{eq:consistent estimator} we can derive that, for every value of $\epsilon$ it exists a value of $k$ such that 
Finally, this implies that, for every value of $\epsilon > 0$, there exists a value $k$ such that $$\left | J_{\pi_k, i} (\beta) - J_{\tilde{\pi}, i}(\beta)\right | < \epsilon$$
This, combined with the fact that $\tilde{\pi}$ satisfies the constraints, gives the desired result.
\end{IEEEproof}

%%%%%%%%%%%%%%%%%%%%%%%%%%%%%%%%%%%%%%%%%%%%%%%%%%%%%%%%%%%%%%%%%%%%%%%%%%%%%%%%%%%%%%%%%%%%%%
\subsection{Proof of \cref{prop:JSSQ safety}}\label{proof:JSSQ}%\label{proof:JSEDk}%\label{proof:lyap}
%%%%%%%%%%%%%%%%%%%%%%%%%%%%%%%%%%%%%%%%%%%%%%%%%%%%%%%%%%%%%%%%%%%%%%%%%%%%%%%%%%%%%%%%%%%%%%

Let \(\xi_i\) denote the desired arrival rate obtained by solving the nonlinear problem \eqref{eq:system for minimizing objective function}.  
We introduce a new policy, referred to as \textit{Randomized JSSQ} ($\pi_{RJSSQ}$ for brevity), which selects the destination server at each timestep based on a probability distribution 
$$\pi_{RJSSQ}(s, i) = \frac{\xi_i}{\sum_j \xi_j}$$
where the $\xi_i$s are determined by solving \eqref{eq:system for minimizing objective function}

\begin{proposition}
    The policy Randomized JSSQ satisfies the constraints.
\end{proposition}
\begin{IEEEproof}
    By definition, under $\pi_{RJSSQ}$ each queue has arrival rate $\xi_i$ solving \eqref{eq:system for minimizing objective function} and therefore satisfies the constraints.  
\end{IEEEproof}

This allows us to prove \cref{prop:JSSQ safety}:
\begin{IEEEproof}
When the occupation of every queue is above their respective threshold $L_i$, 
JSSQ's coincides with Randomized JSSQ. In all the other states, arrivals are assigned only to the set of queues which have an occupation lower than their threshold, according to the proportion implied by their corresponding values of $\xi_i$. 

Hence, JSSQ can be seen as a randomization between two safe policies: Randomized JSSQ, and a safe assignment based on the instantaneous occupation. As such the resulting policy is safe. 
%Hence, for every queue $i$ the average expected occupation is not higher than the one given by Randomized JSSQ and therefore all the constraints are satisfied.
\end{IEEEproof}
\fi

\end{document}